\documentclass[a4paper,11pt]{article}
\usepackage{pos}
\usepackage{amsmath,graphicx,multirow,tabularx,physics}
\usepackage{davitex,slashed}

\title{Streamlined data analysis in Python}
\ShortTitle{AnalysisToolbox}
\author[a]{L. Altenkort}
\author*[b]{D. A. Clarke}
\emailAdd{clarke.davida@gmail.com}
\author[c]{J. Goswami}
\author[a,1]{H. Sandmeyer\note[1]{Work conducted while still at Bielefeld.}}

\affiliation[a]{Universit\"at Bielefeld, Fakult\"at f\"ur Physik,\\
D-33615 Bielefeld, Germany}

\affiliation[b]{Department of Physics and Astronomy, University of Utah,\\
Salt Lake City, Utah  84112, United States}

\affiliation[c]{RIKEN Center for Computational Science,\\
Kobe 650-0047, Japan}

\abstract{
Python is a particularly appealing language to carry out data analysis, owing in part to 
its user-friendly character as well as its access to well maintained and powerful libraries 
like NumPy and SciPy. Still, for the purpose of analyzing data in a lattice QCD context, 
some desirable functionality is missing from these libraries. Moreover, scripting languages 
tend to be slower than compiled ones. To help address these points we present the 
AnalysisToolbox, a collection of Python modules to facilitate lattice QCD data analysis. Some 
highlighted features include general-purpose jackknife and bootstrap routines; modules for 
reading in and storing gauge configurations; a module to carry out hadron resonance gas 
model calculations; and convenience wrappers for SciPy integration, curve fitting, and 
splines. These features are sped up behind the scenes using parallelization and just-in-time
compilation.}

\FullConference{%
 The 40th International Symposium on Lattice Field Theory, LATTICE2023
  31st July - 4th Aug, 2023
  Batavia, Illinois, USA
}

\begin{document}
\maketitle

\section{Motivation}\label{sec:motivate}
Lattice QCD calculations are broadly separated into three distinct steps: one
generates configurations, makes measurements on those configurations, then
carries out data analysis on the measurements. These steps are roughly ranked 
in decreasing order of computational demand. Hence the first two steps usually 
have to be carried out on powerful hardware using high-performance code.
The last step can often be done on one's own laptop. 
There are many general techniques that apply to a large class of statistical problems, 
so it's useful to have code where your implementation of that technique can be 
applied in as many situations as possible. This strategy saves you time; moreover if 
you've implemented the technique correctly, it prevents you from making any error 
having to again implement that technique any further time.

At this stage of analysis it's therefore valuable to use Python. Python has many
characteristics that make it particularly appealing to lattice practitioners
doing data analysis. For instance it
lets you implement classes to keep code well organized; 
doesn't require almost any advanced knowledge of software development; 
has many already existing, well tested libraries\footnote{Nowadays machine
learning is taking a major foothold in lattice field theory, with lattice
practitioners developing algorithms for e.g. gauge field generation and phase transition
identification. Python has access to many machine-learning libraries, and it may
be useful in a lattice field theory context to leverage these.} 
that can be easily imported or exist already in common virtual 
environments such as Anaconda; and 
is very flexible and user friendly.
On the other hand, it's slow relative to compiled languages such as C++ and even
the up-and-coming scripting language, Julia. At the level of data analysis this often 
doesn't matter, but it can, for instance when implementing systematic corrections 
depending on double or triple integrals or solving slowly-converging optimization
problems.

In these proceedings, we present a partial answer to some of the aforementioned
challenges, the {\it AnalysisToolbox}, which is publicly available on 
GitHub~\cite{github}. Statements in these proceedings apply to release 
\ff{v1.1.0}.
Hauke Sandmeyer wrote the basis for this package in the context of HotQCD collaboration. 
From that basis, we 
tried to keep only the most general routines, like those that can jackknife an 
arbitrary function of a time series. In addition, much effort was made to speed up 
such routines, for example through just-in-time compilation or parallelization.

\section{An invitation}\label{sec:invite}

We begin with some examples how the AnalysisToolbox can be used to carry out
physics calculations. 
The hadron resonance gas (HRG) model is a low-temperature statistical physics
model that imagines that hadrons are the only degrees of freedom. Comparing HRG
with lattice data is often utilized to, for example, determine the temperature
at which hadronic states dissociate into quarks~\cite{Sharma:2022ztl}.
In \lstref{lst:HRG} we show how the \ff{HRG} class
can be used to determine the baryon-number fluctuation $\chi_2^B$ in an HRG.
As one can see from the \ff{gen\_chi} call, arbitrary conserved-charge cumulants
are supported.

\begin{code*}
\mycode{python}{code/HRG.py}
\caption{An example HRG calculation. In this case we work at fixed $\mu_B/T=0.3$
and compute $\chi_2^B$ as function of temperature. What is needed is as much
relevant input knowledge about hadron bound states as is known. This is read
from a table in lines 16-17. The table of information
is provided with the AnalysisToolbox. The list is also available as an ancillary 
file in the arXiv version of \cite{Bollweg:2021vqf}.
Line 18 determines for each species whether the gas is bosonic or fermionic.
Finally the \ff{HRG} class is instantiated in line 21, which besides cumulants
like the one computed on line 25,
contains many methods for various thermodynamic observables such as the
pressure and entropy.}
\label{lst:HRG}
\end{code*}

To complete a lattice calculation, one must extrapolate to the continuum limit;
hence straightforward continuum-limit extrapolations are something one will do
fairly often. In the next example, we show how the AnalysisToolbox can be used to
perform general continuum-limit extrapolations.
For instance suppose we want to perform a continuum-limit extrapolation to
determine the deconfinement transition temperature $\Td$ in pure $\SU(3)$. The order
parameter for this phase transition is given by the Polyakov loop, $P$. 
The transition is first-order in the
thermodynamic limit, where $\ev{|P|}$ as function of temperature
would jump discontinuously at $\Td$.
At finite volume, this abrupt jump becomes smooth, and $\Td$ is estimated by the
inflection point of the curve.

\begin{code*}
\mycode{python}{code/continuumExtrapolate.py}
\caption{Example continuum-limit extrapolation. As one sees in line 22, this
calculation was performed on ensembles with aspect ratio 3. Excluding the
imports, we were able to read in $\ev{|P|}$ vs. $T$ tables, fit those data 
with splines,
estimate inflection points, bootstrap those estimates, continuum
extrapolate those results, plot them, and compare the final estimate using a Z-test to the
literature value with about 25 commands.
Taken from a pedagogical project for first- and second-year undergraduate
students~\cite{sri}.}
\label{lst:contExtrap}
\end{code*}

In \lstref{lst:contExtrap} we show how such an extrapolation is achieved with
the AnalysisToolbox, along with error estimation, plotting the results, and
carrying out a statistical comparison with the known literature value.
We assume you already\footnote{Incidentally, the AnalysisToolbox already has code
that will assist with computing some Polyakov-loop observables, such as $|P|$ or
the susceptibility $\chi_{|P|}$, if you started with measurements at the
configuration level.} have results for $\ev{|P|}$ at various $N_\tau$, which we
read in from tables of the form \ff{Nt6.txt}. 
For each $N_\tau$, this code estimates the inflection
point of $\ev{|P|}$ as a function of $T$ to get $\Td(N_\tau)$. 
Temperatures are calculated in MeV using a 2017 parameterization~\cite{Burnier:2017bod} of
the Sommer scale $r_0/a$~\cite{Sommer:1993ce}. This 
procedure is wrapped in a user-defined function\footnote{Within this function is
a \ff{getSpline} method that wraps various SciPy spline methods and
\ff{diff\_deriv}, which is our implementation of a central-difference numerical
derivative.} \ff{getTc}, so that errors in the $\ev{|P|}$ data can be 
conveniently propagated into the error in $\Td(N_\tau)$
using a Gaussian bootstrap\footnote{There is also an ordinary bootstrap, in case
you would rather work with data on the configuration level.}, which is done on
line 41. We stress that this scheme allows the bootstrap to be completely
agnostic to the form and contents of the to-be-bootstrapped function.

Having the \ff{Nts}, \ff{Tds}, and \ff{Tderrs}, we are ready to perform a
continuum-limit extrapolation on line 46. This will perform\footnote{Under
the hood, this tries several different \ff{scipy.optimize} algorithms and
chooses the one with the lowest reduced $\chi^2$. If you would like to try
higher-order fits with few data, we also support Bayesian priors.} an
extrapolation to second order in $a^2$, i.e. $\order{a^4}$, print the fit
results to screen, and create a plot of the extrapolation for you.
The arrays \ff{result} and \ff{result\_err} contain the best fit parameters
along with their errors, with \ff{result[0]} being the continuum-limit $\Td$.
In the last block we compare our result with $\Td r_0$ determined in Francis
{\it et al.}~\cite{Francis:2015lha}. The temperatures calculated in this code implicitly had units of
MeV, hence we use the 2014 result~\cite{HotQCD:2014kol} for $r_0$ in physical units converted to
1/MeV. Finally we call \ff{gaudif} to carry out a Z-test.

\begin{figure}
\centering
\includegraphics[width=0.5\linewidth]{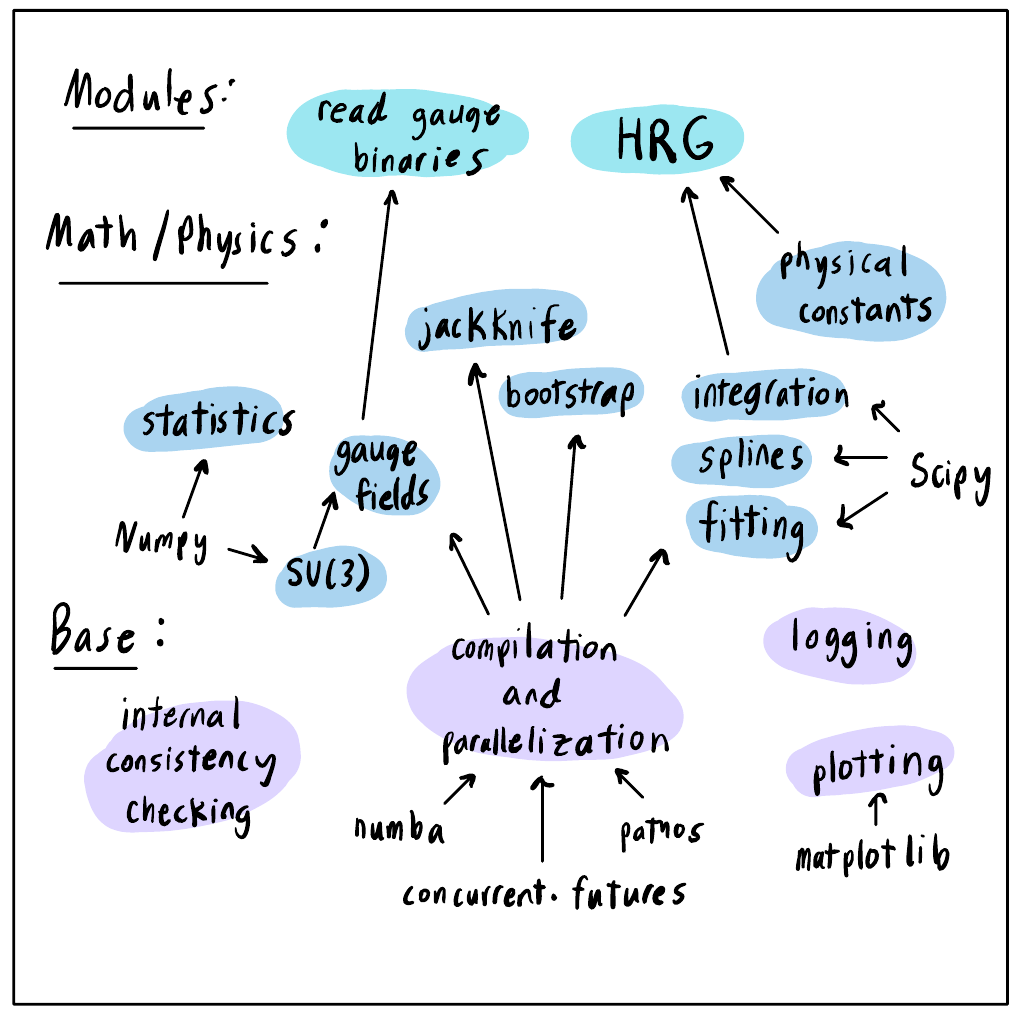}
\caption{Figure 1: Rough sketch of organizational hierarchy. Base modules encapsulate
combinations of well maintained Python modules, such as SciPy and NumPy. 
These are used to construct and enhance math and physics objects, which in turn build 
up high-level modules. Here
the configuration reader and hadron resonance gas model are given as two examples.}
\label{fig:organize}
\end{figure}

\section{Code design}\label{sec:organize}

As mentioned in \secref{sec:motivate}, there exist many well tested and
actively maintained Python packages for various kinds of data analysis. In
general we try to take the philosophy that we cannot write something better than
such libraries, especially given our limited personnel. Therefore, many of the
methods and classes in the AnalysisToolbox are wrappers for or inherit from
existing methods and classes in SciPy~\cite{scipy} and NumPy~\cite{numpy}.

In \figref{fig:organize} we outline the code's organizational hierarchy using
some example modules. At the lowest level we collect methods and classes that
we think all other modules might use. This includes internal methods that
streamline\footnote{For example some methods may expect NumPy arrays, and hence
return an array, even if you wanted to pass a simple scalar. The \ff{unvector}
method checks if something is a single-element, array-like object and, if it is, 
removes the
outermost brackets.} and check for consistency\footnote{For instance checking
whether arrays have the same length or objects are of the correct type. In
particular we try to make sure that, when the code gets something it didn't
expect, it tells you exactly what values it received.}. Error messages are
output with our custom logger, which uses parts of Python's \ff{logging} library.
Our plotting library consists mainly of convenience wrappers for 
Matplotlib~\cite{matplotlib}.

Most importantly at this level is a module for speeding up Python calculations.
Many kinds of computations easily profit from parallelization; for example
a jackknife can be easily parallelized by having each jackknife sample
independently evaluated in parallel. Python has many parallelization libraries
available to it. In our code, we have abstracted this away with
a \ff{parallel\_function\_eval} method. This takes as arguments a function and
an array, then evaluates the function on the array in parallel. Hence it can be
used to replace for loops. While parallelization in Python is generally less
challenging than other languages, it can still be tricky to pass arguments
besides the to-be-parallelized array to the function, and the abstraction helps
with this. Moreover this abstraction lets one switch out different parallelizers
without modifying any higher-level code. 

Sections of Python code using nested C-like for loops are often tremendously slow.
In some instances, they may not be easy to vectorize, and may be slow even after
some straightforward parallelization. For some such cases, Python has available 
to it a just-in-time compilation library called Numba~\cite{numba}. On the other hand,
compilation also takes some time, and it could be that for small enough problem
sizes, the just-in-time compilation time is similar to the run time. Hence we
introduce a \ff{@compile} decorator along with \ff{numbaON()} and
\ff{numbaOFF()} methods to conveniently toggle compilation for large swathes of
code on and off.

At the next level up in the hierarchy we have physics and math modules. When
possible, we try to use already existing methods in NumPy and SciPy, for
instance for curve fitting, integration, matrix multiplication, and very basic
statistics. At the time of writing, numerical differentiation is slated to no
longer be supported by SciPy, and we have not been able to make other
automatic differentiators work well in the code, so we do our own numerical
differentiation using central differences. More advanced statistical tools not
already in NumPy or SciPy include the bootstrap and jackknife. As discussed in
\secref{sec:invite}, the jackknife and bootstrap are agnostic to the problem at
hand, which saves the user from constantly having to reimplement or tailor a
jackknife approach to each new problem. Moreover they call
\ff{parallel\_function\_eval}, so they are sped up under the hood.
Our class for $\SU(3)$ matrices
inherits from NumPy matrices to leverage that already existing functionality,
and we implement a few methods special to $\SU(3)$ like reunitarization.
Parallelization, compilation, and the $\SU(3)$ class are synthesized in 
a gauge field class, so that one can store gauge fields and 
somewhat quickly compute average plaquettes.

At the highest level are modules requiring a synthesis of many lower-lying
modules in the code. The gauge field reader, inspired by \ff{qcdutils}~\cite{qcdutils},
is a good example of this. It reads
in binary and stores the result in a gauge field object. Mirroring gauge field
readers in typical lattice codes, we also check things like the average
plaquette. At the moment we can only read in NERSC format binaries and lack checksum
calculation. 

Finally we mention our unit tests, which are intended to
make sure results remain stable under changes to the code. These tests are
usually very simple and include comparisons against trusted results and
analytic results or comparisons between multiple computation strategies.

\section{Existing features}

\subsection{Math and statistics}

The math and statistics methods of the code are independent of any physics
context and hence are applicable well beyond the lattice QCD.
These methods include
numerical differentiation and integration;
spline fitting;
curve fitting with~\cite{Lepage:2001ym} and without Bayesian priors;
ordinary and Gaussian bootstrap;
jackknife;
autocorrelation calculation~\cite{berg_markov_2004}; and 
the statistical Z-test and T-test.

\subsection{Physics}

We feature physics modules that might be of interest both to lattice
practitioners and those studying finite-temperature QCD phenomenology like
HotQCD parameterizations of, e.g., $af_K(\beta)$ and 
$r_1m_s(\beta)$ \cite{HotQCD:2014kol,Bollweg:2021vqf};
physical parameters and their errors from the PDG, e.g. $m_\pi$ and $m_\rho$; 
the hadron resonance gas model~\cite{Goswami:2020yez,HotQCD:2012fhj};
the QCD equation of state~\cite{Bollweg:2022fqq,Goswami:2022nuu,Bazavov:2017dus};
the QCD beta function;
the static quark potential and Polyakov loop observables; and
critical exponents for various universality classes.

\subsection{Interfacing}

Having evolved in the context of HotQCD and MILC projects, the code interfaces
with some software and conventions of these groups, for instance parsing HotQCD, MILC, and 
SIMULATeQCD~\cite{Altenkort:2021cvg,HotQCD:2023ghu} ensemble names to extract metadata
and jackknifing C. Schmidt's \ff{DenseCode} output~\cite{Allton:2005gk}. We also try to make the code
flexible to conventions in the broader lattice community, 
like
reading\footnote{We have found a gauge field reader to be of use in
Python, since some projects explore applying machine learning at the
configuration level, and many machine learning tools already exist for Python.} 
in NERSC-format gauge configurations and
reading \ff{.gpl} files~\cite{lsqfit}.

\section{Future work and outlook}

We have implemented features in this code as we encountered them in various
lattice projects. Many of these methods are sufficiently general that we believe
they can especially be of use to other lattice practitioners. Some methods, such
as the QCD EoS, should also be useful to phenomenologists, and still others like
the jackknife, bootstrap, and fitting modules are usable for general data
analysis, even outside of a physics context.

As the ILDG progresses~\cite{Karsch:2022tqw}, we intend to dovetail some of our 
efforts with theirs, for instance supporting \ff{QCDml} markup and reading in ILDG 
configurations.

\section*{Acknowledgements}

This work was supported by the Deutsche Forschungsgemeinschaft (DFG, German Research Foundation) 
Proj. No. 315477589-TRR 211 and Proj. No. 460248186 (PUNCH4NFDI).

We have had several other contributors to the AnalysisToolbox code over the
years. In particular we thank
O. Kaczmarek, 
L. Mazur, 
M. Sarkar, 
C. Schmidt, 
H.-T. Shu, and
T. Ueding.

DAC would like to thank his students Kai Ebira, Daeton McClure, Coleman Rohde, and Ed Wright
for producing the data showcased by the example script \ff{main\_continuumExtrapolate.py} and
for their hard work and enthusiasm learning about lattice QCD. He would also like
to thank University of Utah's Science Research Initiative for providing a framework to
give undergraduates some opportunity to try research right away.

\bibliographystyle{JHEP}
\bibliography{bibliography}

\end{document}